\documentclass[aps, prd, twocolumn, showpacs,amsmath,amssymb,superscriptaddress, nofootinbib, showkeys]{revtex4-1}

\usepackage[hyperindex]{hyperref}
\usepackage{amsfonts}
\usepackage{graphicx}
\usepackage{wasysym} 

\newcommand{\mnras}{MNRAS}

\newcommand*\Bell{\ensuremath{\boldsymbol\ell}}

\newcommand{\pert}[1]{{\mbox{\tiny (#1)}}}
\newcommand{\pertd}[2]{{\stackrel{\pert{#1}}#2 \! \! }}
\newcommand{\Newt}{{}}
\newcommand{\Gal}{{\mbox{\tiny G}}}
\renewcommand{\vec}[1]{{\bf #1}}

\def\beq{\begin{eqnarray}}
\def\eeq{\end{eqnarray}}
\def\na{\nabla}

\begin{document}

\author{Davi C. Rodrigues}\email{davi.rodrigues@cosmo-ufes.org}\affiliation{Departamento de F\'{\i}sica, CCE, Universidade Federal do Esp\'{\i}rito Santo, Av. Fernando Ferrari 514, Vit\'{o}ria, ES, 29075-910 Brazil}\author{Sebasti\~{a}o Mauro}\email{sebastiao.filho@ifsuldeminas.edu.br}\affiliation{Departamento de F\'{\i}sica, ICE, Universidade Federal de Juiz de Fora,  Juiz de Fora, MG, 36036-330, Brazil}\affiliation{Instituto Federal de Educa\c{c}\~ao, Ci\^encias e Tecnologia do Sul de Minas - IFSuldeMinas, 37410-000, Tr\^es Cora\c{c}\~oes, Minas Gerais, Brasil.}\author{\'{A}lefe O. F. de Almeida}\email{alefe@cosmo-ufes.org}\affiliation{Departamento de F\'{\i}sica, CCE, Universidade Federal do Esp\'{\i}rito Santo, Av. Fernando Ferrari 514, Vit\'{o}ria, ES, 29075-910 Brazil}

\title{Solar System constraints on  Renormalization Group extended General Relativity: \\
{\normalsize  The PPN and Laplace-Runge-Lenz analyses with the external potential effect}}

\begin{abstract}
	General Relativity extensions based on Renormalization Group effects are motivated by a known physical principle   and constitute a class of extended gravity theories that have some unexplored unique aspects. In this work we develop in detail the Newtonian and post Newtonian limits of a realisation called Renormalization Group extended General Relativity (RGGR). Special attention is taken to the external potential effect, which constitutes a type of screening mechanism typical of RGGR. In the Solar System, RGGR depends on a single dimensionless parameter $\bar \nu_\odot$, and this parameter is such that for $\bar \nu_\odot = 0$ one fully recovers GR in the Solar System. Previously this parameter was constrained to be $|\bar \nu_\odot| \lesssim 10^{-21}$, without considering the external potential effect. Here we show that under a certain approximation RGGR can be cast in a form compatible with the Parametrised Post-Newtonian (PPN) formalism, and we use both the PPN formalism and the Laplace-Runge-Lenz technique to put new bounds on $\bar \nu_\odot$, either considering or not the external potential effect. With the external potential  effect the new bound reads $|\bar \nu_\odot| \lesssim 10^{-16}$. We discuss the possible consequences of this bound to the dark matter abundance in galaxies.
\end{abstract}

\pacs{04.50.Kd, 04.25.Nx, 04.62.+v} 

\keywords{General Relativity extensions, Solar System tests, Quantum fields in curved spacetime}

\maketitle

\section{Introduction }

There are diverse motivations for extending gravity beyond General Relativity \cite[e.g.,][]{Capozziello:2010zz, Capozziello:2011et, Clifton:2011jh}, including: quantum gravity, avoidance of singularities, understanding inflation, theoretical and observational improvements on dark matter, alleviating coincidence issues related to dark energy,   and others. Nonetheless,  these extensions must be capable of explaining the success of General Relativity (GR) in the Solar System. This apparently simple test of gravity is, for many cases, a hard test. 

Among the possible extensions of GR  that have consequences at astrophysical or cosmological scales, we consider here  extensions that are based on the Renormalization Group (RG) framework  applied to gravity \citep{Julve:1978xn, Salam:1978fd, Fradkin:1981iu, Nelson:1982kt,Goldman:1992qs, Shapiro:1999zt,Bonanno:2000ep, Bonanno:2001hi, Reuter:2001ag, Bonanno:2001xi, Bentivegna:2003rr, Reuter:2004nx, Bonanno:2004ki,Niedermaier:2006wt, Shapiro:2009dh,Weinberg:2009wa, Bonanno:2012jy, Sola:2013fka}. In this context, either from Quantum Field Theory (QFT) in curved spacetime \citep{Nelson:1982kt, Shapiro:2008sf}, or in certain quantum-gravity theories, like the asymptotic safety program \citep{Niedermaier:2006wt, Percacci:2007sz, Reuter:2012id}, the GR constants $G$ and $\Lambda$ necessarily run in the ultraviolet limit. Moreover,  in the low energy limit, the $\beta$-functions of $G$ and $\Lambda$ need not to be zero, since the Appelquist-Carazzone decoupling \cite{Appelquist:1974tg}  does not hold in this context (contrary to the coupling constants associated to the high derivative terms that appear in QFT in curved spacetime \cite{Gorbar:2002pw}). Therefore, large scale variations of these `constants' can be a sign of these RG effects, which may as well provide leads to the underlying quantum-gravity theory \cite{Reuter:2004nx}.  Here we will focus on the realization that was named RGGR \citep{Rodrigues:2009vf, Rodrigues:2015hba}.

Besides being motivated from known physical effects, these gravity extensions based on the RG represent a new route to GR extensions on their own. They share  similarities with scalar-tensor gravity in the sense that they depend on a metric and additional quantities that transform as scalars, like $G$, $\Lambda$ or the RG scale $\mu$. These similarities at large scales were particularly explored in Refs.~\cite{Reuter:2003ca, Koch:2010nn, Hindmarsh:2012rc}.\footnote{There are others approaches that can be promptly spotted  as not similar to scalar-tensor theories, in particular those whose RG effects are not implemented at the action level \cite[e.g.,][]{Sola:2013fka}, or those that use additional dynamical tensors \cite[e.g.,][]{Manrique:2010am}, but these are not in the focus of this work.} As detailed in \cite{Rodrigues:2015hba}, which follows and extends the approaches of \cite{Reuter:2003ca, Shapiro:2004ch}, they may also include features  that are either unnatural or clearly outside from the usual scalar-tensor approaches, namely: i) natural and simple potentials from the RG perspective are rather complex from the scalar-tensor perspective, ii) potentials within the RG perspective need not to be universal, they can be derived for each system and can be different from system to system, and iii) the identification of the physical meaning of the RG scale $\mu$ is an important step from the RG perspective, and this identification leads to the imposition of a constraint between $\mu$ and the matter fields. In conclusion, although there are similarities, there are important differences. Also, these differences can lead to new forms of screening mechanisms \cite{Vainshtein:1972sx, Khoury:2003aq, Hinterbichler:2010es, Koivisto:2012za}, and this is one of the main points of this work, namely to show that RGGR has a type of screening mechanism that depends on the external Newtonian potential of a system (in general it depends on the scalar $U^\alpha U^\beta h_{\alpha \beta}$, which will be explained latter).

There are some procedures for testing gravity beyond Newtonian gravity, and the most general and cited one is the Parametrised Post Newtonian (PPN) formalism \cite[e.g.,][]{Nordtvedt:1970uv, Will:1972zz,Blanchet:1989fg,Damour:1990pi,Damour:1991yw,Damour:1992qi, Damour:1993zn, Will:1993ns, Klioner:1999cv,Blanchet:2003gy,Blanchet:2013haa,Will:2014kxa}. Here we consider both a version of the PPN formalism and the Laplace-Runge-Lenz (LRL) vector technique \cite[e.g.,][]{0201657023, 1984AmJPh..52..909S, BenYaacov:2010rc}. 

Solar System tests of RGGR were evaluated in Refs.~\cite{Farina:2011me, Zhao:2015pga}, where bounds on the dimensionless parameter $\bar \nu_\odot \equiv \alpha_\odot \nu$ were found to be respectively $|\bar \nu_\odot| \lesssim 10^{-17}$ and $|\bar \nu_\odot| \lesssim 10^{-21}$. In \cite{Rodrigues:2009vf} it was argued that if the effective constant $\bar \nu$ approximately  runs  linearly with the system mass, then the galaxy results could be explained and also the first bound \cite{Farina:2011me} would be satisfied. If RGGR has significant impact on dark matter  then $\bar \nu$ should be about $\sim 10^{-9} - 10^{-7}$ in galaxies \cite{Rodrigues:2012qm, Rodrigues:2014xka}. Since the baryonic mass of the studied galaxies ranged from $10^8 M_\odot$  to $10^{11} M_\odot$, the first bound is compatible with the linear behaviour, while the second one is not (this was also discussed in \cite{Zhao:2015pga}). The precise mechanism for the running of $\bar \nu$ was unclear.

The previous works on Solar System constraints \cite{Farina:2011me, Zhao:2015pga} have not considered the external potential effect, which is presented in detail here, in Sec \ref{sec:externalpot}. This effect is inherent of the RGGR approach and works as a (partial) screening mechanism. 

In this work, we also clarify the correspondence and the validity of using the RGGR noncovariant approach (introduced in \cite{Rodrigues:2009vf}) as an approximation to the covariant version \cite{Rodrigues:2015hba}. The appendices \ref{app:covariant} and \ref{app:noncovariantASapproximation} are devoted to this issue. 

The starting point of this work is the covariant RGGR formulation proposed in \cite{Rodrigues:2015hba}. Section \ref{sec:review} is devoted to a brief review on RGGR. Section \ref{sec:PNframework} presents a proper post-Newtonian framework, introduces the external potential effect and presents in detail a point-particle solution. Section \ref{sec:PPNeLRL} applies the previous results to the Solar System within the Parametrised Post-Newtonian (PPN) formalism and the Laplace-Runge-Lenz (LRL) technique. This technique can be applied either with or without the external potential effect, hence allowing for an evaluation of the effect relevance. Considering the external potential effect, both techniques can be applied and are compatible. Finally, in Sec.~\ref{sec:conclusions} we present our conclusions and discussions.

\section{Renormalization Group extended General Relativity: a brief review} \label{sec:review}

\subsection{The action}
In \cite{Rodrigues:2015hba} we proposed the following action for describing the large scale RG effects in gravity, which we will in general label as RGGR (Renormalization Group extended General Relativity),
\begin{equation}
    \label{eq:RGNOexterConst}
    S =  \int \left[ \frac{R - 2 \Lambda\{\mu\}}{16 \pi G(\mu)}  + \lambda \left( \mu - f(g, \gamma,\Psi)\right)\right] \sqrt{-g} \,  d^4x + S_{m},
\end{equation}
where $S = S[g,\gamma, \mu,\lambda,\Psi]$, $S_m = S_{m}[g,\Psi]$, $\Psi$ stands for any matter fields of any nature, and $\mu$ is the RG scale, whose relation to all the other fields is stated in the action in a constraint-like way, as imposed by the Lagrange multiplier $\lambda$. The field $\gamma_{\alpha \beta}$, which only appears inside $f$ and without derivatives, is a tensor that works as a reference metric. Reference or background metrics commonly appear in  QFT in curved spacetime and in some quantum gravity approaches. As shown in Ref.~\cite{Rodrigues:2015hba}, $\gamma_{\alpha \beta}$ is  important for guaranteeing energy-momentum conservation, and for presenting certain noncovariant scale settings in covariant form.  About the scalars $G$ and $\Lambda$, they are  respectively the gravitational coupling and the cosmological ``constant'', and they both  depend on the RG scale $\mu$. Namely, $G$ is a standard function of $\mu$, which is fixed at the action level. This means that the form of this dependence is independent on the other fields and their boundary conditions (i.e., if, for instance, $G = \mu^2$ for cosmology, then $G = \mu^2$ for the Solar System, for all the galaxies, for vacuum and for any other system). On the other hand, the relation between $\Lambda$ and $\mu$ is not assumed to be universal, it is system-dependent. It is not fixed at the action level, but it can and must be derived from the field equations. This is why we introduced in \cite{Rodrigues:2015hba}  different notations for these dependences, we write $G(\mu)$ and $\Lambda \{ \mu \}$.

From the RG perspective, the difference between $G(\mu)$ and $\Lambda \{ \mu \}$ is that for the first one we assume the existence of a universal $\beta$-function, that is, a $\beta$-function that is independent on any other properties of the system; while for $\Lambda \{ \mu \}$ the corresponding $\beta$-function is system dependent. Since $\beta$-functions in general depend on the presence of other fields (and gravity interacts with everything), the use of $``\{ \mu \}"$ should not come as a surprise. Since there are diverse works that suggest a simple and specific form for $G(\mu)$, for the latter only we use the usual fixed dependence $``(\mu)"$. Whenever it is necessary to specify a function $G(\mu)$ we use the following  simple expression  that has been  derived from different approaches \cite[e.g.,][]{Fradkin:1981iu, Nelson:1982kt, Reuter:2003ca,  Shapiro:2004ch, Bauer:2005rpa},
\begin{equation}
    \label{eq:Gmu}
    G^{-1}(\mu) = 1 + 2 \nu \ln \mu,
\end{equation}
where $\nu$ is a small dimensionless constant, and GR is recovered for $\nu=0$.

\subsection{A class of covariant scale settings and $T^{\alpha \beta}$ conservation}

The action (\ref{eq:RGNOexterConst}) in general spoil the energy-momentum tensor conservation, but  there is a particular class of scale settings (i.e., $f$ functions)  that preserves energy-momentum conservation, namely \cite{Rodrigues:2015hba},
\begin{equation}
    \label{eq:mufUUh}
    \mu = f\left( { U^\alpha U^\beta h_{\alpha \beta}} \right),
\end{equation}
where $h_{\alpha \beta} \equiv  g_{\alpha \beta} - \gamma_{\alpha \beta}$.  The tensor $U^\alpha$ denotes the four-velocity field. The coupling  of the RGGR action to a general perfect fluid is presented in Ref.~\cite{Rodrigues:2015hba}, which uses the fluid action description of Ref.~\cite{1972JMP....13.1451R}.

On the energy-momentum conservation,  the action (\ref{eq:RGNOexterConst}) does not lead in general to $\nabla_\alpha T^{\alpha \beta}=0$, where $T^{\alpha \beta}$ is the matter energy-momentum tensor. The reason being that it is not possible to write $S = S_g[g,\Phi_\Gal] + S_m[g,\Psi]$, where $\Phi_\Gal$ denotes fields of any nature that do not appear in $S_m$, while $\Psi$ denotes fields that do not appear in $S_g$ (for a review, see Ref.~\cite[][Appendix E]{Wald:1984rg} ). Since $\lambda$ is the term that prevents the action splitting into gravitational and matter parts, in general $\nabla_\alpha T^{\alpha \beta}\propto \lambda$. Hence, if $\lambda$ is set to zero at the level of the field equations,  it is possible to guarantee that $\nabla_\alpha T^{\alpha \beta}=0$. This is precisely achieved by using the scale setting proposed in eq. (\ref{eq:mufUUh}). Indeed, since $\gamma_{\alpha \beta}$ only appears in the action inside $f$, the variation of $S$ with respect to $\gamma_{\alpha \beta}$ leads to
\begin{equation}
	\lambda f' U^\alpha U^\beta	= 0. \label{eq:lambdazero}
\end{equation}
Hence either  $\lambda =0$ or $f'=0$. The last option simply implies that $\mu$ is a constant, leading to standard GR. A GR extension can be found if the other solution holds, namely if $\lambda=0$ at the level of the field equations, which  implies that $\nabla_\alpha T^{\alpha \beta}=0$.

For the scale setting (\ref{eq:mufUUh}), the variation of the action (\ref{eq:RGNOexterConst}) with respect to the metric reads
\begin{equation}
    \label{eq:fieldext}
    {\cal G}_{\alpha \beta} + \Lambda g_{\alpha \beta} = {8 \pi G} T_{\alpha \beta},
\end{equation}
where 
\begin{equation}
        {\cal G}_{\alpha \beta} \equiv G_{\alpha \beta} +  g_{\alpha \beta} G \Box G^{-1}  - G \nabla_\alpha \nabla_\beta G^{-1},
\end{equation}
$\Box \equiv g^{\alpha \beta} \nabla_\alpha \nabla_\beta$, and $\nabla_\alpha$ is the usual covariant derivative. From the energy-momentum tensor conservation, one derives that \cite{Koch:2010nn, Rodrigues:2012qm, Rodrigues:2015hba}
\begin{equation}
    \label{eq:consist}
    \nabla_\alpha \left( \frac{\Lambda}G \right) = \frac 12 R \nabla_\alpha G^{-1}. 
\end{equation}
The equation above can also be derived from the action variation with respect to $\mu$. Further details on the RGGR action can be found in Ref.~\cite{Rodrigues:2015hba}.

\subsection{The noncovariant scale setting}

An issue that any RG approach to gravity must answer is the physical meaning of the scale $\mu$,  that is relation of $\mu$ with other physical quantities (which in the end is the same of specifying the $f$ function). 

In the context of stationary, slow velocity and weak field systems, some of us have introduced in previous works the scale setting \cite{Rodrigues:2009vf}
\begin{equation}
    \label{eq:muRGGR}
    \mu =  \left( \frac{\Phi_\Newt}{\Phi_0} \right)^\alpha,
\end{equation}
where $\Phi_0$ and $\alpha$ are constants that describe the system, $\Phi_\Newt$ is the Newtonian potential, defined by
\begin{equation}
	\nabla^2 \Phi_\Newt = 4 \pi G_0 \rho \, ,  \mbox{ with } 	\Phi_\Newt(r \rightarrow \infty) = 0. \label{eq:PhiN}
\end{equation} 
In the above,  $\rho$ is the matter density, $G_0$ is the gravitational constant at some spacetime point, and $\mu$ is the RG scale written in dimensionless form. In the following, a  system of units such that $G_0 = 1$ is always used. The constant $\Phi_0$ is actually irrelevant in any perturbative expansion up to the first order of  $\nu$.

The noncovariant scale setting above can be seen as an approximation to the covariant one, as detailed in the Appendix \ref{app:covariant}.

The scale setting (\ref{eq:muRGGR}) has achieved interesting phenomenological consequences for galaxy systems. In particular we considered its implications to dark matter \cite{Rodrigues:2009vf, Rodrigues:2011cq, Fabris:2012wg, Rodrigues:2012qm, Rodrigues:2012wk, Rodrigues:2014xka, deOliveira:2015cja}, and it was found that this approach can have a significant impact on the necessary amount of dark matter in galaxies. Indeed, the internal dynamics of galaxies alone shows that good results are achievable in galaxies even without dark matter.

As previously stated, $\alpha$ is not a  universal constant (contrary to $\nu$), it depends on the system. Considering galaxy rotation curves (e.g., \cite{Rodrigues:2009vf, Rodrigues:2014xka}), it is used as a constant inside a galaxy, but it changes from galaxy to galaxy.\footnote{This behaviour can be qualitatively described by the covariant formulation, as detailed  in the Appendices. Nonetheless there is no known covariant expression that can quantitatively explain these variations} Since all the dynamical tests depend not on $\alpha$ alone, but on the combination $\alpha \nu$, effectively one can replace the two constants $\alpha$ and $\nu$ by  a single system dependent constant $\bar \nu \equiv \nu \alpha$. In \cite{Farina:2011me} it was found that, for the Solar System internal dynamics, $|\bar \nu_\odot|  \lesssim 10^{-17}$. More recently, Ref. \cite{Zhao:2015pga} used  more precise data for the Solar System and arrived at the condition  $|\bar \nu_\odot| \lesssim 10^{-21}$. Nevertheless, neither of these references considered the external potential effect, which is detailed in Sec. \ref{sec:externalpot}.

\section{Post-Newtonian framework, the external potential effect and point particle solution} \label{sec:PNframework}

This section is devoted to three itens not covered in previous publications: i) introducing a perturbative scheme that will allow for the application of the PPN formalism, ii) to present the external potential effect, which is a kind of screening mechanism that is part of RGGR, and iii) to present a detailed evaluation of the point particle solution from the field equations such that it can be used in the PPN and LRL analyses in the next section.

Although all the results present in this section are derived from the noncovariant scale setting, the same results can also be derived from the covariant one, as shown in the appendices \ref{app:covariant} and \ref{app:noncovariantASapproximation}.

\subsection{GR and RG perturbations}\label{sec:GR RG pert}

Consider the following perturbative scheme about the metric $\stackrel{\pert{0}}g_{\alpha \beta}$,
\begin{eqnarray}
	   g_{\alpha \beta} &=& \stackrel{\pert{0}}g_{\alpha \beta} +  \pertd{1,0}{g}_{\alpha \beta} +   \pertd{0,1}g_{\alpha \beta} + ..., \label{eq:perturbRGg2}\\[.1in]
    T_{\alpha \beta} &=& \stackrel{\pert{0}}T_{\alpha \beta} + \pertd{1,0}T_{\alpha \beta} +   \pertd{0,1}T_{\alpha \beta} + ..., \label{eq:perturbRGT2} \\[.1in]
	G(\mu) &\equiv&  1 + \delta G (\mu) = 1 +  \pertd{1}G\; (\mu) +...\label{eq:perturbRGG2} \\[.1in]
    \Lambda(\mu) &\equiv& \Lambda_0 + \delta \Lambda(\mu) = \Lambda_0 + \pertd{1}\Lambda \; (\mu) +... \label{eq:perturbRGL2}
\end{eqnarray}
The metric $\stackrel{\pert{0}}g_{\alpha \beta}$ satisfies the Einstein equation with the energy momentum tensor $\stackrel{\pert{0}}T_{\alpha \beta}$, the gravitational constant $G_0$ (which is set to be 1) and the cosmological constant $\Lambda_0$.  The terms of the type $\pertd{n,0}X_{\alpha \beta}$ refer  to some perturbative expansion within  GR; for instance, $n$ may refer to the order of a post-Newtonian expansion.  The terms of the type $\pertd{n,m}X_{\alpha \beta}$ are the  RG correction of $m$-th  order to the GR perturbation of order $n$. 

The background is here picked to be Minkowski, that is,
\begin{equation}
		\stackrel{\pert{0}}g_{\alpha \beta} = \eta_{\alpha \beta},  \;\;\;\;\;  \Lambda_0 =0, \;\;\;\;\;  \pertd{0}T_{\alpha \beta} =0.
\end{equation}

Within this case, it was shown in detail in \cite{Rodrigues:2015hba} that, up to first order on both of the perturbations, if $\tilde g_{\alpha \beta}$ is a solution of the Einstein equation given by $\tilde G_\alpha^\beta = {8 \pi} \tilde T_\alpha^ \beta$, then the metric solution for the field equation (\ref{eq:fieldext}) can be found from the conformal transformation
\begin{equation} \label{eq:gGtildeg}
	g_{\alpha \beta} = G \,  \tilde g_{\alpha \beta} + O(2,2).
\end{equation}
The symbol $O(m,n)$ designates any terms of the m-th or higher order on the GR perturbation, and of n-th or higher order on the RG perturbation. For the particular case of $O(2,2)$, when it is present it is implied that the mixed terms on  the perturbations are not explicitly written (since the terms of the type $\pertd{1,1}X$ are necessary equal or smaller than either $\pertd{2,0}X$ or $\pertd{0,2}X$). The use of   $O(\infty,m)$, which will appear latter, implies that the expression is exact if $\nu =0$, it is an exact GR expression with RG corrections up to the order $m-1$.

In order to illustrate the notation and review an important result that can also be found in Refs.~\cite{Rodrigues:2009vf, Rodrigues:2015hba}, let $\tilde g_{\alpha \beta} = \eta_{\alpha \beta} + \tilde h_{\alpha \beta}$ and $ g_{\alpha \beta} = \eta_{\alpha \beta} +  h_{\alpha \beta}$, therefore, using eq.~(\ref{eq:gGtildeg}),
\begin{eqnarray}
		h_{\alpha \beta} &=&  g_{\alpha \beta} - \eta_{\alpha \beta} \nonumber \\[.1in]
		&=&  (\eta_{\alpha \beta} + \tilde h_{\alpha \beta}) G - \eta_{\alpha \beta} + O(2,2)\label{eq:hhtilde}.	
\end{eqnarray}
Hence, in particular,
\begin{eqnarray}
	h_{00} &=& - 2 \Phi_\Newt 	- \pertd{1}G\; + O(2,2) \nonumber \\[.1in]
	&=& - 2 \Phi_\Newt 	+ 2  \nu \ln \mu   + O(2,2). \label{eq:effectivepotentialrggrO22}
\end{eqnarray}

Since $h_{00}$ is twice the effective potential (i.e., the potential whose gradient yields the acceleration), the above equation expresses the relation between the RGGR effective potential and the Newtonian potential (up to first order on both the perturbations).

\subsection{The external potential effect} \label{sec:externalpot}

Here we consider the dynamical effect of an external Newtonian potential. It is shown that the larger is the external potential absolute value, the smaller are the non-Newtonian effects of the considered system. Hence, an external potential  acts as a screening mechanism for RGGR, in the sense that the environment reduces the RGGR non-Newtonian (and non-GR) contribution. To be clear, this effect is not an {\it ad-hoc} feature, it is already part of the theory.\\

\noindent
{\bf The spherically symmetric case.} With the scale setting (\ref{eq:muRGGR}) or (\ref{eq:muUUhsimple}), which sets a relation between $\mu$ and $\Phi_\Newt$, gravity is in general sensitive  to the Newtonian potential value, such that the dynamics of a system may change due to a constant shift on the Newtonian potential. 

In particular, for a static system with spherical symmetry, it is possible to define an effective additional mass of RGGR ($\delta M_{\mbox{\tiny RGGR}}$) which  can be expressed as \cite{Rodrigues:2012qm}
\begin{eqnarray}
		\delta M_{\mbox{\tiny RGGR}} (r) &\equiv & (\Phi'_{\mbox{\tiny RGGR}} - \Phi')  {r^2} \nonumber \\[.1in]
		 &=&{\bar \nu } \frac{r}{1 + \frac{4 \pi r }{M(r)} \int_{r}^{\infty} \rho(a) a \, da},
	\label{MRGGR}
\end{eqnarray}
where $\Phi_{\mbox{\tiny RGGR}}$ is the effective potential of RGGR (i.e., its gradient yields the acceleration of a test particle). If the mass distribution is simply that of a particle of mass $M$, the term  with the integral is zero for $r>0$, and  the effective additional mass increases linearly with $r$ (and hence the additional force decreases with $r$). Contrary to Newtonian gravity or pure General Relativity, this result is sensitive to the existence of a spherical mass distribution at large radius. If there is such mass distribution, the integral $ \int_{r}^{\infty} \rho(a) a \, da$ will be greater than zero, and hence $\delta M_{\mbox{\tiny RGGR}}$ will be suppressed. 

In conclusion, considering a static spherical mass distribution, the larger is the amount of mass outside a given region, the smaller are the non-Newtonian effects in that region. Equation (\ref{MRGGR}) re-expressed the external potential effect as an external mass density effect, with the hypothesis that stationary spherical symmetry holds.\\

\noindent
{\bf Average potentials and subsystems.}
Let $S'$ be a subsystem of a system $S$, much smaller than $S$, such that at $S'$ the average Newtonian potential of $S$ (denoted by $\Phi_{\Newt s}$) is a constant. The use of the term ``average'' is to be explicit that structures whose characteristic size are much smaller than the size of $S$ are not individually considered. Similarly, the (average) Newtonian potential of a galaxy does not consider individual stars, and the (average) Newtonian potential inside of a star does not consider individual particles. Hence the Newtonian potentials of $S$ and $S'$ can be written as,
\begin{eqnarray}
	\Phi_{\Newt s} & = & \phi_s + \phi_e \, , \\[.1in]
	\Phi_{\Newt s'} & = & \phi_{s'} + \Phi_{\Newt s}|_{s'}  \nonumber \\ 
	        &=& \phi_{s'} + \phi_s|_{s'} + \phi_e  \nonumber \\
	        &=& \phi_{s'} + \phi_{e'}  + \phi_e \, .
\end{eqnarray}
In the above, $\phi_s$  stands for the Newtonian potential generated by the system $S$, and $\phi_e$ refers to the total external contribution.  The universe is nor static or stationary at large scales, hence $\phi_e$ is {\it a priori} an unknown effective constant.

Since $S'$ is a small subsystem of $S$, all the external contribution is  $\Phi_{\Newt s}|_{s'}$, which is the total (average) Newtonian potential of the system $S$ at the position of the system $S'$. The external potential contribution to the system $S'$ can be described by two terms. One of them is the same $\phi_e$ constant that appears in $\Phi_s$, and the other is $\phi_{e'} = \phi_{s}|_{s'}$.\\

\noindent
{\bf Consequences for $G$ at the system and subsystem.} The expression for $G$ associated to the system $S$, with explicit reference to the constants $G_0$ and $\Phi_0$, reads (using eqs. \ref{eq:Gmu}, \ref{eq:muRGGR})
\begin{equation}
	G_{s}^{-1}(\phi_s)  = G_0^{-1}\left( 1 + 2 \bar \nu_{s} \ln \frac{\phi_{s} + \phi_e}{\Phi_{0}} \right).
\end{equation}

It is also possible to express the same function $G_{s}^{-1}$ with respect to other reference potential. In particular, using $\phi_e$ as the reference potential, one finds,
\begin{equation}
	G_{s}^{-1}(\phi_s)  = G_e^{-1}\left[ 1 + 2 \bar \nu_{e} \ln \left( 1 + \frac{\phi_{s}}{\phi_{e}} \right)  \right],
\end{equation}
with $G_e$ and $\bar \nu_e$ such that
\begin{eqnarray}
		G^{-1}_e \bar \nu_e &=& G^{-1}_0 \bar \nu_s,  \label{eq:GeNuG0Nu}\\[.1in]
		G_e^{-1} &=& G_0^{-1} \left ( 1 + 2 \bar \nu_s \ln\frac {\phi_e}{\Phi_0} \right).
\end{eqnarray}
The above equations are found by demanding that the two expressions for $G_s^{-1}(\phi_s)$ above are compatible among themselves and that the relation between $G_0$, $G_e$, $\Phi_0$ and $\phi_e$ must be a constant, that is, that it cannot depend on $\phi_s$.

Using the relations just presented for changing the reference potential, the $G_{s'}$ function corresponding to subsystem $S'$ can be written as
\begin{eqnarray}
		G_{s'}^{-1}(\phi_{s'})  &=& G_{0'}^{-1}\left( 1 + 2 \bar \nu_{s'} \ln \frac{\phi_{s'} + \phi_{e'} + \phi_e}{\Phi_{0'}} \right) \nonumber \\[.1in]
		&=& G_{e'}^{-1}\left[ 1 + 2 \bar \nu_{e'} \ln \left( 1 + \frac{\phi_{s'} + \phi_e}{\phi_{e'}}\right)  \right].
\end{eqnarray}

If the  unknown constant $\phi_{e}$ satisfies $|\phi_e| \ll  |\phi_s|$ in a given region (which implies that $|\phi_e| \ll  |\phi_{e'}|$), then in this region one can write
\begin{eqnarray}
	G_{s}^{-1}(\phi_s)  &\approx & G_e^{-1}\left( 1 + 2 \bar \nu_{e} \ln \frac{\phi_{s}}{\phi_{e}} \right),\\[.1in]
	G_{s'}^{-1}(\phi_{s'}) &\approx &	G_{e'}^{-1}\left[ 1 + 2 \bar \nu_{e'} \ln \left( 1 + \frac{\phi_{s'}}{\phi_{e'}}\right)  \right] \nonumber \\[.1in]
	&\approx & G_{e'}^{-1}\left[ 1 + 2 \bar \nu_{e'}  \left( \frac{\phi_{s'}}{\phi_{e'}} - \frac 12 \frac{\phi_{s'}^2}{\phi_{e'}^2}\right)  \right]. \label{eq:GslinhaExpansion}
\end{eqnarray}
In the above we used that $\ln(1 + X) = \ln X + O(1/X)$, for $|X| \gg 1$, and $\ln(1 + x) = x - x^2/2 + O(x^3)$, for $|x| \ll 1$. The last expansion assumes $\phi_{s'}/\phi_{e'} < 1$, which is a condition realised in the Solar System context, as it will be shown.

The expansion (\ref{eq:GslinhaExpansion}) will prove useful for the PPN application. Henceforth, for simplicity, we will always consider $|\phi_e| \ll |\phi_s|$. Also, this condition is necessary in order to find compatibility with the galaxy results of Ref.~\cite{Rodrigues:2009vf}.

\bigskip

\noindent
{\bf The Solar System as a subsystem of the Galaxy.} For this application, the system $S$ will be the Milky Way, designated by the symbol ``$_{\mbox{\tiny MW}}$'', and the subsystem $S'$ will be the Solar System, which is designated by ``$_\odot$''. From Refs. \cite{Iocco:2015xga, Pato:2015tja}, one arrives at the following estimates for the value of $\phi_{\mbox{\tiny MW}}|_{\odot}$, that is, the Newtonian potential generated by the Milky Way evaluated at the Solar System position: $- 5 \times 10^{-7}$, if only  the baryonic matter is considered, or $-2.1 \times 10^{-6}$ considering both the baryonic and a standard dark matter halo (both in unities of $c^2$, which is set to be one). Since the RG effects may, at least in part, mimic  dark matter-like effects, the true value of $\phi_{\mbox{\tiny MW}}|_{\odot}$ should lie in between these two cases.

\begin{figure}[hbt]
  \includegraphics[width=0.9\columnwidth]{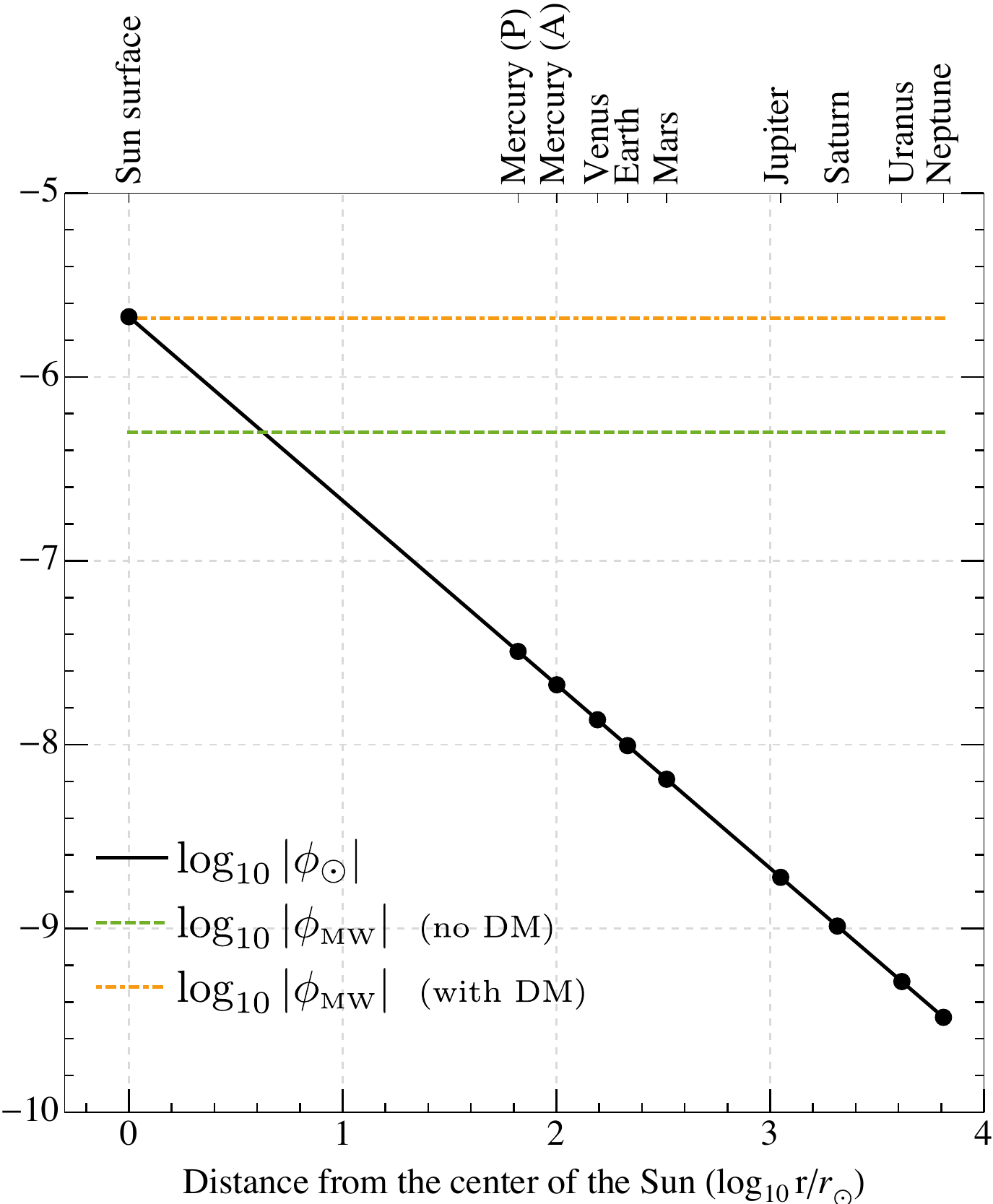}
  \caption{The Newtonian potential generated by the Sun $\phi_\odot$ across the Solar System, and the value of the Newtonian potentials generated by the Galaxy  at the Solar System ($\phi_{\mbox{\tiny MW}}|_{\odot}$).  The letters ``P'' and ``A'' after Mercury refers to its perihelion and aphelion.}
  \label{fig:PlanetsP}
\end{figure}

  Figure  \ref{fig:PlanetsP} shows the Newtonian potential generated by the Sun ($\phi_\odot$) across the Solar System and the values of $\phi_{\mbox{\tiny MW}}|_{\odot}$ either with or without dark matter. The Solar System data was derived from \cite{nasaplanets}. 
  
  The Solar System is a subsystem of the Milky Way, hence its external  potential  $\phi_{e'}$ is  $\phi_{\mbox{\tiny MW}}$. For all the planets one finds $\phi_\odot/\phi_{e'} \lesssim 10^{-2}$, which shows that the expansion (\ref{eq:GslinhaExpansion}) can be used in this context. To be more precise, the largest value of $\phi_\odot/\phi_{e'}$ comes from Mercury at its perihelion, and it reads: $ 6.4 \times 10^{-2}$ for the case without dark matter, and $ 1.5 \times 10^{-2}$ for the case with dark matter.

\subsection{Point particle solution of RGGR: detailed derivation from the field equations} \label{sec:pointparticle}

The purpose of this subsection is to present a derivation of the point particle solution directly from the field equations (\ref{eq:fieldext}). This is not the first time that point particle solutions are considered in the RGGR context, see for instance \cite{Rodrigues:2009vf} and eq.~(\ref{eq:effectivepotentialrggrO22}), where it was  used a conformal transformation method. However, beyond being useful for verifying the result without the use of conformal transformations (whose employment in gravity theories has led to diverse debates on its meaning), the latter method only holds for the first order perturbation in both the GR and the RG expansions, while from the direct use of the field equations it is possible to find analytical results valid up to the order $O(\infty, 2)$, that is, up to arbitrary order on the GR parameters, and apart from second order corrections on the RG parameter $\bar \nu$. 

Considering the proper spacetime symmetry, and without loss of generality in this context, let 
\begin{equation}
ds^2 =  g_{00}(r) dt^2 + g_{11}(r) dr^2 + r^2 d\Omega^2 ,
\label{lineelement}
\end{equation}
with   $d\Omega^2 =  d\theta^2 + \sin^2(\theta) \, d\phi^2$. 

With the above line element, it is straightforward to show 
the following identities,
\begin{eqnarray}
&&{\cal G}_\mu^\nu = 0 \; \; \; \; \forall \mu \not= \nu, 
\nonumber
\\
&& {\cal G}_2^2 = {\cal G}_3^3.
\end{eqnarray}

It is useful to use two field equations, from (\ref{eq:fieldext}), rearranged such that $\Lambda$ does not appear explicitly, that is, 
\begin{eqnarray}
	{\cal G}_0^0 & = & {\cal G}_1^1\, , \label{eq:calG00}	\\[.1in]
	{\cal G}_1^1 & = & {\cal G}_2^2 \, \label{eq:calG11}.
\end{eqnarray}
The remaining nontrivial field equation simply presents the solution for $\Lambda$, and it can be written as $\Lambda = - {\cal G}_0^0$. This approach was also used in \cite{Rodrigues:2012qm, Rodrigues:2015rya}. From the first one, the relation between $g_{11}$ and $g_{00}$ up to first order on $\bar \nu$ is derived to be
\begin{equation}
g_{11} = - \frac{K}{g_{00}} \left[ 1 + 2 \bar \nu    \left(  r \frac{ \mu'}{\mu} - \ln \frac{\mu}{\mu_1}  \right) \right] + O(\infty, 2), \label{eq:g11particle}
\end{equation}
where a prime means derivative with respect to $r$ and $K$ and $\mu_1$ are  integration constants. Within GR, the constant $K$  can be trivially eliminated by a time redefinition ($t \rightarrow t/\sqrt{K}$). Within RGGR, both the constants $K$ and $\mu_1$ can  be absorbed trough a time redefinition. To prove this, one only need to note that
\begin{eqnarray}
	&& K \left (1 + 2 \bar \nu \left (r \frac{\mu'}{\mu} - \ln \frac{\mu}{\mu_1} \right) \right) \approx \nonumber \\	
	 &&\approx	K (1 +  2 \bar \nu  \ln \mu_1) \left(1 + 2 \bar \nu \left (r \frac{\mu'}{\mu} - \ln \mu \right)\right),
\end{eqnarray}
up to first order on $\bar \nu$. And hence, after the time redefinition $t \rightarrow t/\sqrt{K(1 + 2 
\bar \nu \ln \mu_1)}$, the $\mu_1$ constant is eliminated. Equivalently, it can be fixed to be 1.

From the eqs. (\ref{eq:calG11}, \ref{eq:g11particle}), the  $g_{00}$ solution is derived to be
\begin{equation}\label{eq:g00particle}
g_{00} = -1 + \frac{C_1}{r} + C_2 r^2 + 2 \bar \nu \left( 1 - \frac {3 C_1}{2 r} \right) 
\ln\mu  + O(\infty, 2).
\end{equation}
The constants $C_1$ and $C_2$ within General Relativity ($\bar \nu =0$) are associated respectively to a mass at $r=0$ and to  a cosmological constant. As expected, the derived solution is an extension of the Schwarszchild-de Sitter solution. The above result, in the context of a point particle, extends the solution (\ref{eq:effectivepotentialrggrO22}).


\section{The PPN and LRL analyses for the Solar System} \label{sec:PPNeLRL}

\subsection{PPN with the external potential effect}

Reviews on the PPN formalism can be found on diverse references \cite[e.g.,][]{Will:1993ns, Fujii:2003pa, Blanchet:2013haa, Will:2014kxa, 1107032865}. The full PPN approach, as used in \cite{Will:1993ns}, depends on 10 parameters, but scalar-tensor theories have only two parameters whose values can be different from the corresponding GR values, which are commonly denoted by $\gamma$ and $\beta$. As analysed in detail in \cite{Rodrigues:2015hba}, RGGR can be seen as a peculiar type of scalar-tensor theory with certain constraint and with a potential that is system dependent. Since in a given system the potential is fixed, at the Solar System RGGR is expected to have only two nontrivial PPN parameters. To find their values, we consider the line element (\ref{lineelement}) and the following metric expansion \cite{Will:1993ns, Fujii:2003pa},
\begin{eqnarray}
	g_{00} &\approx & 	-1 + \frac{2 M}r - 2 (\beta - \gamma) \left( \frac M r \right )^2 , \label{eq:g00edd}\\[.1in]
	g_{11} &\approx & 1 + 2 \gamma \frac{M}r. \label{eq:g11edd}
\end{eqnarray}
In the above, $\gamma$ and $\beta$ are constants, and for GR both of them are equal to 1. A full detailed analyses of RGGR directly from its covariant expression and without the application restriction to the Solar System is beyond the scope of this work, and is currently a work in progress. This simpler PPN application, sometimes also referred as Eddington expansion, is nonetheless both useful for the Solar System application and to introduce procedures that will be useful for the full PPN development in this context.

Within the PPN framework, it is not uncommon to associate an order of smallness such that $v^n \sim (M/r)^{n/2} \sim O(n)$, where $\pmb v$ is the test particle velocity. Nevertheless, for clarity in the present context, we  use the convention in which $n$ is associated with the power on the metric perturbation, hence we use $(M/r)^{n} \sim O(n)$.

Comparing eqs. (\ref{eq:g11particle}, \ref{eq:g00particle}) with eqs. (\ref{eq:g00edd}, \ref{eq:g11edd}), one sees that a minimum condition for applying this parametrization is $|C_2|r^2 \approx 0$ for the range of $r$ considered and $C_1/r \sim O(1)$. With these considerations, eqs.~(\ref{eq:g11particle}, \ref{eq:g00particle}) can be written as
\begin{eqnarray}
	g_{00} &=& 	-1 + \frac{C_1}r  + 2 \bar \nu \left( 1 - \frac{3 C_1 }{2r} \right )\ln \mu + \nonumber \\ && + O(\infty,2) \,,  \\[.1in]
	g_{11} &=& 1 + \frac{C_1}r + 2 \bar \nu \left(  r \frac{\mu'}{\mu} + C_1 \left(  \frac{\mu'}{\mu} - \frac{\ln \mu}{2r}\right) \right) + \nonumber \\ & & + O(2,2)\, . 
\end{eqnarray}
In the above, since $C_2$ is no longer considered, $O(n,m)$ refers to terms of $n$-th order or higher on $C_1$, and terms of $m$-th order or higher on $\bar \nu$. All the terms that depend on  $C_1 \bar \nu$ are considered.

The ``$\ln$'' terms can be expanded as in eq. (\ref{eq:GslinhaExpansion}). It is convenient to write explicitly the dependence on $r$, hence let
\begin{equation}
	\frac{\phi_\odot}{\phi_{e'}} \equiv \frac k r\,.	 \label{eq:k}
\end{equation}
With these considerations,
\begin{eqnarray}
	g_{00} &=& -1 + \frac{C_1}r +  \frac{k}{r} \bar \nu \left( -2  - \frac{k}{r} - \frac{2 k^2}{3 r^2}+ 3 \frac{C_1}{r}   + \right. \nonumber \\   && \left. + \frac{3 k C_1 }{2 r^2} + \frac{k^2 C_1}{r^3} \right) + O(\infty,2,4),\label{g00particle2} \\[.1in]
	g_{11} &=& 	1 + \frac{C_1}r  + 2 \frac{k}{r}  \bar \nu \left(1 + \frac{3 C_1}{2 r} \right ) +\nonumber \\ & & + O(2,2,2), \label{g11particle2} 
\end{eqnarray}
where $O(\infty,2,4)$ refer to terms of arbitrary order on $C_1$, of second or higher order on $\bar \nu$ and  of fourth or higher order on $k$. The meaning of $O(2,2,2)$ follows analogously.  There is no terms with $C_1^{\, 2}$ in $g_{00}$ since there is none of such terms up to first order on $\bar \nu$ (see eq. \ref{eq:g00particle}). 

To apply the formalism, it is necessary to relate the  expansions used above, otherwise it is impossible to know whether, say, $C_1^2/r^2$ can be neglected while and $k^2 \bar \nu /r^2$ is considered. For the case of the planet Mercury, the terms of order $O(1)$ are those of the same order of $\phi_\odot(r_{\mbox{\tiny \mercury}}) = -2.7 \times 10^{-8}$ (the Newtonian potential generated by the Sun at Mercury's orbit). This should correspond, apart from higher order corrections, to $-M/r$. In other words, this is the assumption that, at Newtonian level in the Solar System,  RGGR must agree with GR and Newtonian theory, otherwise there is no hope to be compatible with the Solar System data. Moreover, $\bar \nu$ needs to be sufficiently small. Considering galaxy internal dynamics, the upper bound found for $\bar \nu$ was $|\bar \nu| \lesssim 10^{-7}$ \cite{Rodrigues:2009vf}, which will be used as a starting point, but soon a stronger bound will be shown.

The value of $k/r_{\mbox{\tiny \mercury}}$ can be computed from its definition (\ref{eq:k}), and it corresponds to $O(0.17)$ for the case without dark matter and $O(0.25)$ for the case with standard dark matter. 

 With the above analysis, the relation between the three expansions is clarified, and it is possible to sort the expansion terms. Hence, eqs. (\ref{g00particle2}, \ref{g11particle2}) can be expressed as
\begin{eqnarray}
	g_{00} &\approx & -1 + \frac{C_1}{r} - \frac{k}r \bar \nu \left(2 + \frac{k}{r} \right ) \, ,	 \label{eq:g00particle3}\\[.1in]
	g_{11} &\approx & 1 + \frac{C_1}r + 2 \frac{k}r \bar \nu \, .\label{eq:g11particle3}
\end{eqnarray}

The relation between $C_1$ and $M$ is fixed by comparing eqs.  (\ref{eq:g00edd}, \ref{eq:g00particle3}), and it yields,
\begin{equation}
	C_1 = 2 M + 2 \bar \nu k \,.
\end{equation}

From the coefficient of $r^{-1}$ in $g_{11}$, $\gamma$ is found to be
\begin{equation}
	\gamma = 1 + \frac{2 \bar \nu k}M = 1 - \frac{2 \bar \nu}{\phi_{e'}} \, .
\end{equation} 

According to Ref.~\cite{Will:2014kxa}, $\gamma $  is constrained from Solar System experiments and observations to satisfy $|\gamma - 1| \lesssim 10^{-5}$. Since $|\phi_{e'}| \sim 10^{-6}$, we derive from the above that $|\bar \nu| \lesssim 10^{-11} $.

From the coefficient of $r^{-2}$ in eq. (\ref{eq:g00particle3}), the value of $\gamma$ and eq. (\ref{eq:g00edd}), $\beta$ can be derived as
\begin{equation}
	\beta = 1 + \frac{\bar \nu k^2}{2 M^2} = 1 + \frac{\bar \nu }{2 \phi_{e'}^{\, 2}}.
\end{equation}
Since $|\beta - 1| \lesssim 10^{-4}$ \cite{Will:2014kxa}, the above implies that 
\begin{equation}
	|\bar \nu_\odot| \lesssim 10^{-16}.
\end{equation}

  We stress that the external potential effect is essential for the PPN application in its standard form \cite{Will:1993ns}, and that this bound above considers it. In the next subsection, by using the LRL approach,  this bound with the external potential effect is confirmed, and it will be compared to the case without it.

\subsection{LRL vector dynamics with and without the external potential effect} \label{sec:lrl}

This section is devoted to estimate the upper bound on $\bar\nu$ in the Solar System from the Laplace-Runge-Lenz (LRL) vector dynamics using the perihelion precession data from \cite{Iorio:2014roa}, which is more recent and precise than the data set used in \cite{Farina:2011me}.

Reviews on the LRL vector can be found in \cite{0201657023, 1984AmJPh..52..909S, BenYaacov:2010rc}. The notation and approach used here follow closely those of Ref.~\cite{Farina:2011me}.

One of the most important results of general relativity is predicting a correction to the precession of the orbit of the planets. New determinations of the corrections to the usual Newtonian-Einsteinian secular precession of perihelion of the planets constitute a relevant data set to constraint modified gravitation models in the Solar System, see Table \ref{tab:estimated}.

\begin{table}[h]
\centering
\caption{Estimated corrections, in milliarcseconds per century,
	     to the standard Newtonian-Einsteinian secular precessions of the
	     perihelion determined with the INPOP10a and the EPM2011
	     ephemerides \cite{Fienga:2011qh, 2013MNRAS.432.3431P}. The relevant data for the purposes of this work are the uncertainties in this table.}
\begin{tabular}{l|lll}

Planet   & EPM2011                & INPOP10a \\
\hline
Mercury  &  -2.0$\pm$3.0          & 0.4$\pm$ 0.6\\
Venus    &  2.6$\pm$ 1.6          &  0.2 $\pm$  1.5    \\
Earth    & 0.19$\pm$ 0.19         &  -0.20 $\pm$  0.90    \\
Mars     &-0.020$\pm$ 0.037       &  -0.040 $\pm$ 0.150     \\
Jupiter  & 58.7$\pm$ 28.3         &  -41.0 $\pm$  42.0    \\
Saturn   & -0.32$\pm$ 0.47        & 0.15$\pm$ 0.65 \\
\end{tabular} \label{tab:estimated}
\end{table}

From  eq.~(\ref{eq:effectivepotentialrggrO22}),  the RGGR gravitational potential, apart from the terms $O(\nu^2)$ and $O(\nu \Phi)$, reads
\beq 
\Phi_{\mbox{\tiny RGGR}} =  \Phi - \nu  \ln\mu \ . \label{eq:PhiRGGRLRL}
\eeq
The above potential is essentially the one that appears in  eq.~(\ref{eq:g00particle3}) and was used for PPN, but without using the $k$ expansion.  Although the PPN approach started from a more precise framework, with computations valid to arbitrary order on $C_1$, in the end there was no significant change on the RGGR potential in comparison with the one derived from the conformal transformation. It should be stressed, however, that the LRL technique only tests the planet orbits, while the PPN approach used here tests  both the orbits (from $\beta$) and the light deflection due to the Sun (from $\gamma$). Therefore, since the major RGGR constraint found from the PPN formalism came from  the $\beta$ observational constraints, the LRL analyses should yield essentially the same bound on $\bar \nu_\odot$. This will be confirmed in this section.

From eq.~(\ref{eq:PhiRGGRLRL}), for a point particle of mass $m$ the force is given by 
\beq
\vec{F}_{\mbox{\tiny RGGR}} = \vec{F_N} +  m \nu \vec{\na}\ln\mu \ ,
\eeq
where $\vec{F_N}$ is the Newtonian gravitational force for a point particle.
Using the effective $\mu$ from eq. (\ref{eq:muRGGR}), and considering the external potential effect, 
\beq
\vec{F}_{\mbox{\tiny RGGR}} = \vec{F_N} -  \frac{m\bar\nu r_0 }{r(r+r_0)} \,\vec{\hat{r}} \label{new-force} \ ,
\eeq
where $r_0=-M_\odot /\phi_{e'}$ and $\vec{\hat r}$ is the standard unit radial vector with origin at the Sun. The above is the gravitational force acting on a point particle of mass $m$ in the weak field regime. 

The LRL vector associated to the Sun with mass $M_\odot$ and a planet of mass $m$ is given by 
\begin{equation}
	\vec{A} = \vec{p} \times \Bell - m^2 M_\odot \vec{\hat r}, 
\end{equation}
where $\vec{p}$ is the linear momentum of the particle of mass $m$, $\Bell=\vec{r} \times \vec{p}$ is the angular momentum. Some important properties of this vector are that $\vec{A} \cdot \Bell =0$ and that, for an unperturbed Newtonian gravity, $d \vec{A}/dt = \vec{0}$ and  $\vec{A}$ is colinear to the major axis. In case of any perturbation, in general $\vec{A}$ will not be a constant of motion and will slowly precess. Also, the magnitude of the LRL vector yields a relation to the eccentricity $\varepsilon$,
\begin{equation}
	|\vec{A}| = m^2 M_\odot \varepsilon.
\end{equation}

The average precession of the orbit is derived from the computation of $\langle d \vec A /dt \rangle$ and reads \cite{Farina:2011me}
\beq
\vec{\Omega} = - \frac{\langle F_p \cos \varphi \rangle}{M_\odot m^2 \varepsilon}  \Bell  \, ,
\label{vel-precession1}
\eeq
where $\varphi$ is the angle between $\vec{A}$ and the major semiaxis, the symbol $\langle \;\; \rangle$ means an average in the following sense,
\begin{equation}
	\langle X \rangle \equiv \frac{m}{\ell \tau} \int_0^{2\pi} r^2(\varphi) \, X(r(\varphi), \varphi) \, d\varphi,
\end{equation}
and $\tau$ is the period of the unperturbed motion.

From eq. (\ref{vel-precession1}), the orbit precession velocity  associated to the force (\ref{new-force}) reads,
\beq
{\Omega} \,=\,
\dfrac{2 \pi \bar\nu \,a(1-\varepsilon^2)}
{M_\odot \tau \varepsilon^2} \left[\frac{1}{\sqrt{1 - \big[\frac{r_0\epsilon}{r_0+a(1-\varepsilon^2)}\big]^2}} - 1 \right]  ,
\label{vel-precession2}
\eeq
where $a$ is the major semiaxis of the ellipse. The above generalizes the $\Omega$ expression for RGGR of Ref.~\cite{Farina:2011me}, by considering the presence of an external potential. Namely, the expression of Ref.~\cite{Farina:2011me} is found in the limit $r_0 \rightarrow \infty$.

\begin{table}
	\centering
	\caption{Orbital parameters of the planets. Here, $\tau$ is the orbital period,	$\varepsilon$ is the eccentricity and $a$ is the major semiaxis of the orbit \cite{nasaplanets}.} 
	\begin{tabular}{l|lll}
		Planet   & $\tau$(years) &\,\,\,\, $\varepsilon$ & $a$ ($10^{10}$ m) \\
		\hline
		Mercury  &    0.241   & 0.2056    &   5.791        \\
		Venus    &   0.615    &  0.0067   &   10.82        \\
		Earth    &    1       &   0.0167  &   14.96        \\
		Mars     &   1.881    &  0.0935   &   22.792       \\
		Jupiter  &  11.862    & 0.0489    &  77.857        \\
		Saturn   &  29.457    &  0.0565   &  143.353       \\
	\end{tabular} \label{tab:parameters}
\end{table}

Using eq. (\ref{vel-precession2}) with the values of the perihelion precession of the planets, as given by Table \ref{tab:estimated}, together with the values of $a$, $\tau$ and $\varepsilon$ from Table \ref{tab:parameters}, one can find an upper bound on $\bar\nu$ for each of the planet orbits. The $\bar \nu$ bounds  with and without the external potential can be seen respectively in Table \ref{tab:LRL}.

\begin{table}
\centering
\caption{Upper bound on $|\bar\nu_\odot|$ either with or without the external potential effect (EPE), using $\phi_{e'} \sim 10^{-6}$. The data for $\Omega$ came from the uncertainties in Table \ref{tab:estimated}. Among the two samples in that table,  the smallest uncertainties were selected for each planet.}
\begin{tabular}{l| c c}
		
Planet   & $|\bar\nu_\odot|$ {\footnotesize (with EPE)} & $|\bar\nu_\odot|$ {\footnotesize (without EPE)}                \\
\hline
Mercury  &    $\lesssim$ $10^{-16}$   &   $\lesssim$ $10^{-19}$    \\
Venus    &   $\lesssim$  $10^{-15}$    &  $\lesssim$  $10^{-19}$  \\
Earth    &    $\lesssim$  $10^{-16}$   &  $\lesssim$  $10^{-20}$ \\
Mars     &   $\lesssim$  $10^{-16}$    &   $\lesssim$  $10^{-20}$  \\
Jupiter  &  $\lesssim$   $10^{-12}$    &  $\lesssim$   $10^{-17}$  \\
Saturn   &  $\lesssim$   $10^{-13}$    &   $\lesssim$   $10^{-19}$\\

\end{tabular} \label{tab:LRL}
\end{table}

\section{Conclusions} \label{sec:conclusions}

Currently there are research lines that look for General Relativity (GR) extensions based on the Renormalization Group (RG) flow of $G$ and $\Lambda$, and consider their possible effects to large scale (infrared) physics. These extend  GR by using principles that are well stablished in other contexts. The approaches that are natural within this RG framework are either impossible or unnatural to achieve by other means, thus it introduces new paths for extending and evaluating gravity. Also, finding nontrivial flows of $G$ and $\Lambda$ in the large scales may provide clues on quantum-gravity \cite{ Reuter:2004nx, Niedermaier:2006wt}.

In this work we focused on a particular realisation named RGGR \cite{Rodrigues:2009vf, Rodrigues:2015hba}. It is based and extends the approaches of \cite{Reuter:2003ca, Shapiro:2004ch}. This extension depends on an effect dimensionless quantity $\bar \nu$ that measures the strength of the RG in a given system, and it is such that in the limit $\bar \nu \rightarrow 0$ the theory becomes pure GR. Two previous works have found bounds on $\bar \nu$ at the Solar System ($\bar \nu_\odot$) \cite{Farina:2011me, Zhao:2015pga}, but they have not considered the external potential effect, which we present here for the first time. This effect is part of RGGR either within the noncovariant or the covariant formulations, and it acts as a new kind of screening mechanism.

Considering this external potential effect, we found that $|\bar \nu_\odot| \lesssim 10^{-16}$ either from the PPN formalism or the LRL vector technique. The external  potential effect could alleviate the bounds associated to the Solar System planet orbits from three to six orders of magnitude (see Table \ref{tab:LRL}). However, this effect alone cannot fully explain the difference between the effective $\bar \nu$ in the Solar System from that in a galaxy, in case RGGR does have a significant impact on galaxy dark matter. 

The external potential effect  acts as a screening mechanism for RGGR in the sense that, the larger is the external potential, the smaller are the non-GR corrections. In other words, the environment can in principle hide the RG effects. Quantitatively, the external potential effect is not sufficient to completely hide the RG effects in the Solar System, if the RG effects are relevant for galactic dynamics; but, in the end, it is a significant dynamical effect that should always be considered.

The change of $\bar \nu$ from system to system may follow a linear correlation to the system mass, as argued in \cite{Rodrigues:2009vf, Rodrigues:2014xka}, and also be compatible with the bounds here derived for the Solar System. The appendices develop further on the  effective changes of $\bar \nu$ from system to system, but the precise mechanism that may allow for a variation of $\bar \nu$ of about eight to ten orders of magnitude from the Solar System to a galaxy, if there is one, it is still unclear.

Independently on the possible connection to dark matter, here we have evaluated the RGGR Solar System bounds,  introduced in detail the external potential effect for the first time, developed an approach for applying standard PPN formalism to  RGGR (this approach requires the use of of the external potential effect), and used the LRL technique to evaluate the magnitude of the external potential effect. This latter effect may open new possibilities on screening mechanisms, not necessarily related to the RGGR approach.

\begin{acknowledgments}
We thank Tomi Koivisto, David Mota and Ilya Shapiro for commenting on previous versions of this work, and Jos\'e de Freitas Pacheco for a discussion on the PPN parameters. SM and AOFA thank CAPES (Brazil) for support. DCR thanks CNPq (Brazil) and FAPES (Brazil) for partial financial support.  	
\end{acknowledgments}

\appendix

\section{A specific covariant scale setting} \label{app:covariant}

Consider the following simple realization of the covariant setting (\ref{eq:mufUUh}),
\begin{equation}
	\mu = f(U^\alpha U^\beta h_{\alpha \beta}) =  A + B \, U^\alpha U^\beta h_{\alpha \beta} , \label{eq:muUUhsimple}
\end{equation}
where $A$ and $B$ are constants. This simple covariant scale setting was introduced in \cite{Rodrigues:2015hba}, and it will be shown in detail in this and the next appendix that, under certain reasonable limits, it is as a covariant extension of the scale setting (\ref{eq:muRGGR}). 

Adopting a comoving coordinate system ($U^i=0$), the scalar $U^\alpha U^\beta h_{\alpha \beta}$ can be expressed as, with $\gamma_{\alpha \beta} = \eta_{\alpha \beta}$, 
\begin{eqnarray}
	U^\alpha U^\beta h_{\alpha \beta} &=& U^0U^0h_{00} \nonumber \\
	&=& - \frac{h_{00}}{g_{00}} \nonumber \\
	&=& - \left( 1 + \frac{1}{G \bar g_{00}} \right ) \label{eq:U0U0h00}\\
	&=& -1 + \frac{G^{-1}}{1 -  \bar h_{00}}. \nonumber
\end{eqnarray}
In the above, it was used $\bar g_{00}  \equiv G^{-1}  g_{00}$. The above expression fixes a relation between $\mu$ and $\bar h_{00}$. For a Minkowski background, following Section \ref{sec:GR RG pert},  the relation between $\bar h_{00}$ and the Newtonian potential $\Phi$, reads
\begin{equation} \label{eq:barh00}
	\bar h_{00} = - 2 \Phi + O(2,2),
\end{equation}
where it was used that the metric that solves the Einstein equation $\tilde G_{\alpha \beta} = 8 \pi \tilde T_{\alpha \beta}$ is $\tilde g_{\alpha \beta}$, whose time-time component satisfies $\tilde g_{00} = -1 - 2 \Phi + O(2)$, and that $\tilde g_{00} = G^{-1}g_{00} + O(2,2) =  \bar g_{00} + O(2,2)$.

The function $\mu(\bar h_{00})$, or $\mu(\Phi)$, will not be an analytical function in general. Indeed, considering the $G(\mu)$ expression as given in eq.~(\ref{eq:Gmu}),  the equation (\ref{eq:U0U0h00}) is a transcendental one for $\mu$.

Far away from any mass, $h_{\alpha \beta}$ should become zero (i.e., the metric $g_{\alpha \beta}$ should coincide with the background), hence  in this limit $\mu = A$, which in turn implies that $G^{-1} = 1 + 2 \nu \ln A$. Using  unities such that  $G|_{h_{\alpha \beta} = 0} = G_0 = 1$, one finds 
\begin{equation}
	A = 1.	
\end{equation}
To  avoid any singularity in $G$ for any $\mu \in [1,\infty)$, $\nu$ needs to be positive, and this is always assumed henceforth.

Combining  the previous equations, 
\begin{eqnarray} 
	\mu &=& 1 + B \left( -1 + \frac{G^{-1}}{1 - \bar h_{00}} \right)  \\[.1in]
	&=& 1 + B \left ( -1 + \frac{1 + 2 \nu \ln \mu}{1 + 2 \Phi} + O(2,2) \right) . \nonumber
\end{eqnarray}
This is a transcendental equation for $\mu$, but it can be  solved for $\Phi_\Newt$, 
\begin{equation}
	\Phi = \frac{1}{2}\left( \frac{1 + 2 \nu \ln (1 + \delta \mu) }{1 + \frac{\delta \mu}{B}} -1 \right) + O(\nu^2).	\label{eq:A6}
\end{equation}
In the above,  we introduced $\delta \mu \equiv \mu - 1 > 0$. As expected, from the above one finds  $\lim_{\delta \mu \rightarrow 0}\Phi_\Newt = 0$.

Up to this point, $B$ is simply any real number, but from the previous results, and two considerations, its value can be  found. For sufficiently small $\delta \mu $, $\Phi$ reads
\begin{equation}
			\Phi|_{\mbox{\tiny $\begin{matrix} \delta \mu \ll |B| \cr \delta \mu \ll 1 \end{matrix}$}}  \approx  \frac 12 \left (  2 \nu  \delta \mu  - \frac{\delta \mu}{B}\right) = \frac{1}2 \delta \mu (2 \nu - B^{-1}). \label{eq:Philimit1}	
\end{equation}
The first consideration is that the inequality $\Phi \leq 0 $ must be satisfied, hence, since $\nu>0$, 
\begin{equation}
	0<B \leq \frac 1{2\nu}.
\end{equation}
The second consideration is that when $\delta \mu \rightarrow 0$ or equivalently when $h_{\alpha \beta} \rightarrow 0$, $\Phi$ should smoothly go to zero, implying that
\begin{equation} \label{eq:PhiDelmu}
	\lim_{\delta \mu \rightarrow 0}\partial_{\delta \mu }\Phi=0.
\end{equation}
Therefore, 
\begin{equation}
	B = \frac 1 {2 \nu}.
\end{equation}
With the above, eq.~(\ref{eq:A6}) can now be simply written as
\begin{equation} \label{eq:A6plus}
	\Phi =  \nu \ln (1+\delta\mu) - \nu \delta \mu + O(\nu^2).	
\end{equation}
To clarify the meaning of the above equation, it is stating a correlation between $\delta \mu$  and $\Phi$, and this correlation, naturally, only exists if $\nu \not=0$. This correlation is the one that comes from the covariant scale setting, and should be compared with the noncovariant one (\ref{eq:muRGGR}). The above equation cannot be solved analytically to express either $\delta \mu$ or $\mu$ as a function of $\Phi$, hence eq.~(\ref{eq:muRGGR}) should be seen as a local analytical approximation for the  function $\mu(\Phi)$.

\begin{figure}[hbt]
  \includegraphics[width=0.95\columnwidth]{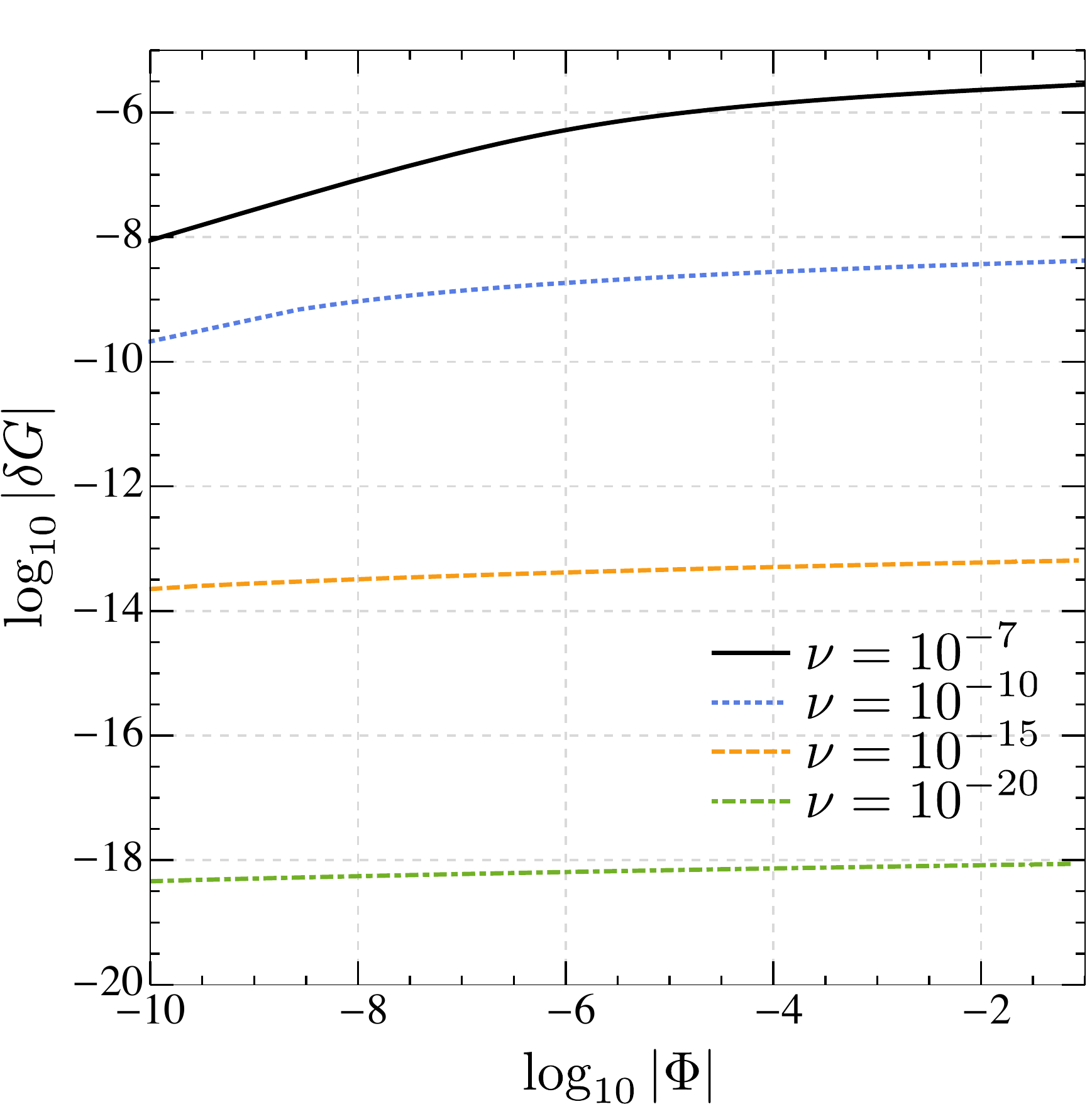}
  \caption{The relation between $|\delta G| = |G - 1| \approx  2 \nu \ln \mu $ (from eq. \ref{eq:Gmu}) and $|\Phi_\Newt|$ (from eq. \ref{eq:A6}) for four different values of $\nu$.}
  \label{fig:plotPhiG}
\end{figure}

Figure \ref{fig:plotPhiG} shows a parametric plot on the evolution of $\delta G$ as a function of $\Phi$ for different values of $\nu$. The highest value of $\nu$ used in that figure corresponds to the value used in galaxies without dark matter \citep[e.g.,][]{Rodrigues:2012qm, Rodrigues:2014xka}, and the smallest one is close to the Solar System bound derived in \cite{Zhao:2015pga}. It can be seen that changes of many orders of magnitude on $\Phi$ translate into a much smaller variation in $\delta G$. The range of $\Phi$ includes values corresponding the surface of a neutron star ($\sim 10^{-1}$), and down to $10^{-10}$, which is about the Newtonian potential generated by the baryonic matter of dwarf galaxies at their farthest observed rotation curve radius.

\section{The noncovariant scale setting as an approximation for the covariant one} \label{app:noncovariantASapproximation}

\begin{figure}[h]
 \includegraphics[width=0.95\columnwidth]{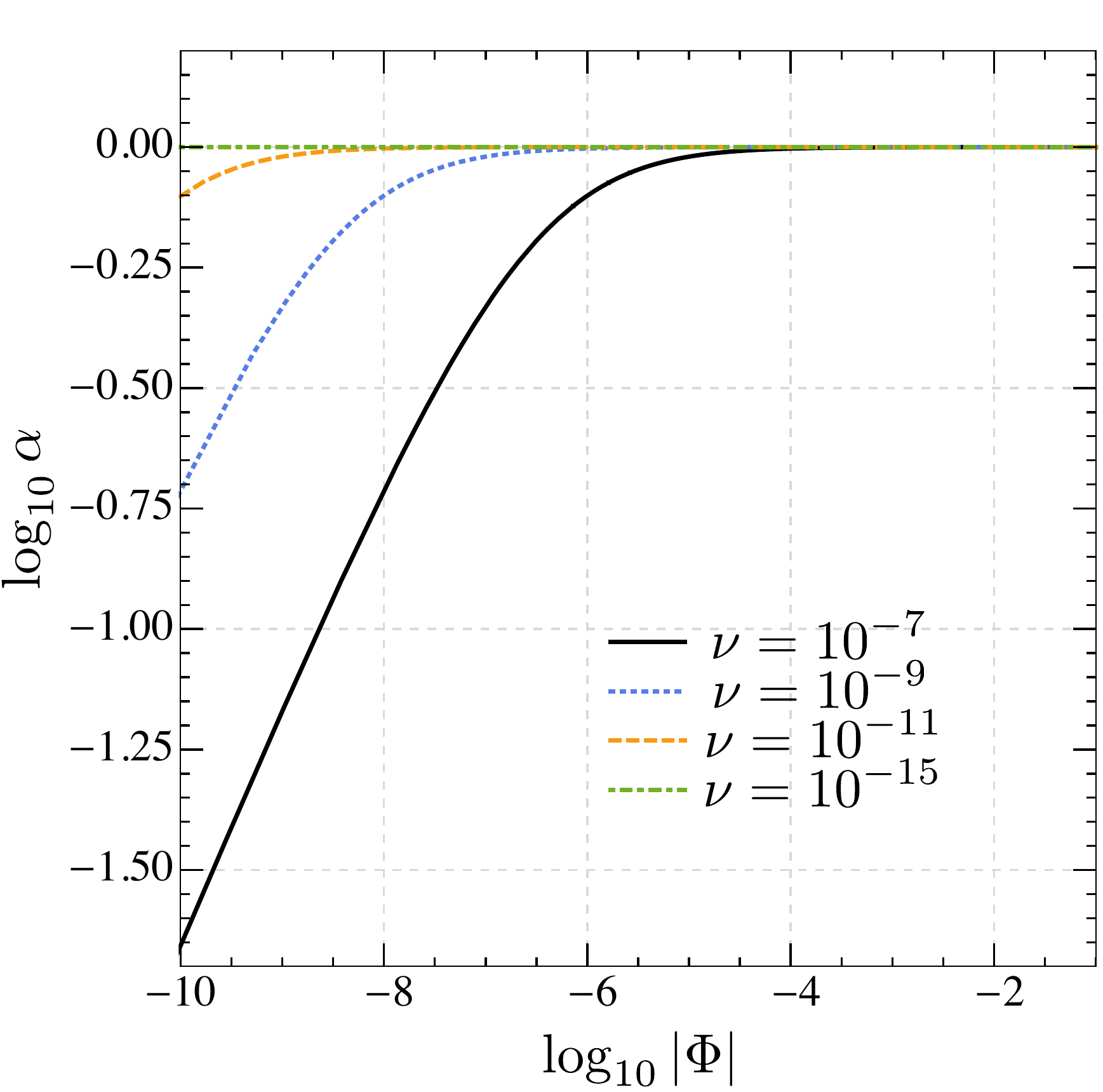}
  \caption{The relation between $\alpha$ and $\Phi$ for different values of $\nu$. It shows that the noncovariant approximation, where $\alpha$ is a constant, can be a good approximation for many systems, since large changes of $\Phi$ lead to much smaller changes on $\alpha$. }
  \label{fig:plotAlphaPhi}
\end{figure}

Equation (\ref{eq:muRGGR}) implies that
\begin{equation}
	\alpha = \frac{1}{\mu \, \partial_\mu \ln \left ( - \Phi_\Newt \right )}, \label{eq:alphamu}
\end{equation}
where $\partial_\mu$ is the derivative with respect to $\mu$.   In eq.~(\ref{eq:muRGGR}), $\alpha$ appears as a constant, but from the perspective of the covariant scale setting, $\alpha$ should in general be a function of $\mu$, as given by the above equation. If, for a given system, $\alpha$ is close to a constant, then for that system the noncovariant scale setting may work as a good approximation. The relation between $\alpha$ and $\Phi$ can be seen in Fig.~\ref{fig:plotAlphaPhi}, which indeed shows that $\alpha$ changes slowly even if $\Phi$ changes by some orders of magnitude.

By using the noncovariant approach, one is using an approximation to derive the PPN parameters of the covariant approach. This approximation must be sufficiently precise. One way to evaluate this error is to consider, from a given value of $\Phi$, the relative error between the two $\mu$'s inferred from the eqs. (\ref{eq:muRGGR}, \ref{eq:A6plus}). Since it is possible to analytically express $\Phi(\mu)$, it is more convenient to adopt the inverse route, that is, from a given $\mu$, to find the relative error between the potentials inferred by   eqs.~(\ref{eq:muRGGR}, \ref{eq:A6plus}), which we call here $\Phi$ and $\Phi_A$ respectively. If the maximum relative error between $\Phi$ and $\Phi_A$, along the Mercury's orbit, is $\varepsilon_\nu^{\mbox{\tiny max}}$, then the PPN parameter $\gamma$, and consequently $\beta$, acquires an additional uncertainty of $\pm  \gamma \varepsilon_\nu^{\mbox{\tiny max}}$ when inferred from the other approach. In particular, if $\varepsilon_\nu^{\mbox{\tiny max}} \sim 10$, it is no longer possible to use one approach (the noncovariant scale setting) to  state precisely the order of magnitude of either $\gamma$ or $\beta$ of the other approach (the covariant scale setting case).  

In order to show that both the approaches lead to compatible bounds for Mercury's orbit, we  compare the Newtonian potential from eq.~(\ref{eq:A6plus}), $\Phi(\mu)$, to an approximated potential given by eq.~(\ref{eq:muRGGR}), namely
\begin{equation}
\Phi_A(\mu, \mu_0) \equiv \mu^{1/\alpha(\mu_0)} \Phi_0(\mu_0),	
\end{equation}
where $\alpha(\mu)$ is given by eq.~(\ref{eq:alphamu}), and $\Phi_0(\mu_0)$ is such that $\Phi_A(\mu_0,\mu_0) = \Phi(\mu_0)$. In the plot of Fig.~\ref{fig:plotAlphaPhi}, the above approximation corresponds to a straight line approximation at $\mu_0$ to the $\alpha(\mu)$ curve. To quantify the approximation, we use the relative error that is given by 
\begin{equation} \label{eq:varepsilon}
	\varepsilon_\nu(\mu,\mu_0) \equiv \left | 1 - \frac{\Phi_{\Newt}(\mu)}{\Phi_{A}(\mu,\mu_0)} \right|.
\end{equation}

Without considering the external potential effect, the range of $\Phi$ values of relevance is from $  - 3.2 \times 10^{-8}$ to  $- 2.1 \times 10^{-8}$. The contribution of the Milky Way to the local potential depends on whether dark matter is being considered or not, but for both cases it is about $\phi_{\mbox{\tiny MW}}(r_\odot) \sim 10^{-6}$ at the Solar System position. This means that the range of variation of the Newtonian potential along the orbit of Mercury is $ [- (K_{\mbox{\tiny MW}} + 0.032) \times 10^{-6}, -(K_{\mbox{\tiny MW}} + 0.021) \times 10^{-6} ]$, where $K_{\mbox{\tiny MW}}$ is a number about the unity whose precise value depends on the amount of dark matter in the Milky Way. 

\begin{figure}[h]
 \includegraphics[width=0.95\columnwidth]{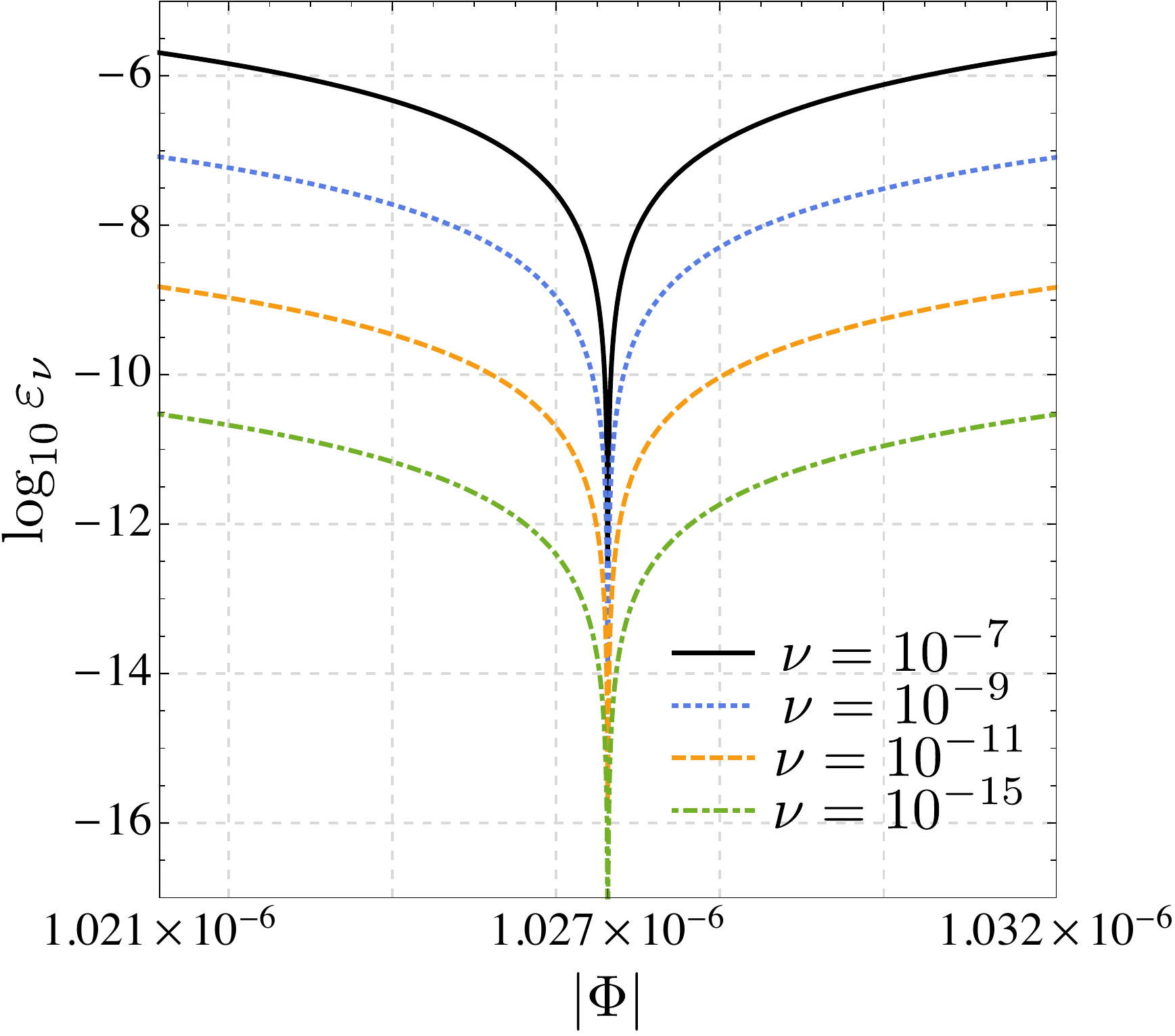}
  \caption{The relative errors (\ref{eq:varepsilon}) introduced by the use of the approximation (\ref{eq:muRGGR}),  in the context of Mercury's orbit with the external potential of the Galaxy. The range of $\Phi_\Newt$ in this plot spans the variation of the Newtonian potential through the planet orbit added by $10^{-6}$. Dividing or multiplying the latter value by  5, does not change the value of $\log_{10} \varepsilon$  significantly. The plot indicates that the noncovariant scale setting works as a good approximation for the covariant one, in the context of  Mercury's  orbit.}
  \label{fig:errornu}
\end{figure}

Figure \ref{fig:errornu} shows that the relative error introduced by the approximation (\ref{eq:muRGGR}) is small enough to allow for a PPN evaluation for the planet Mercury for all the relevant values of $\nu$. For the case without external potential effect, one can draw a similar plot as that of Fig. \ref{fig:errornu}, with higher values of the relative errors, but no higher than $10^{-2}$.

Since the main focus here is on order of magnitude evaluations of the Solar System bounds, the above shows that there exists a value for $\alpha$ such that eq.~(\ref{eq:muRGGR}) can work as a satisfactory approximation to the covariant scale setting (\ref{eq:muUUhsimple}), considering the post-Newtonian analysis of Mercury's orbit.

\bibliographystyle{apsrev4-1} 
\bibliography{bibdavi2016c}{} 

\end{document}